\documentclass{iau}
\usepackage{graphicx}
\usepackage[]{natbib}

\title[Blazars in radio and gamma-rays] %% [give here short title] %%
{The connection between the 15 GHz radio and gamma-ray emission in blazars}

\author[W. Max-Moerbeck et al.]   %% [give here the short author list; use "et al." if 3 authors or more] %%
{
W. Max-Moerbeck$^1$,
J. L. Richards$^2$,
T. Hovatta$^3$,
V. Pavlidou$^4$,\\
T. J. Pearson$^5$,
A. C. S. Readhead$^5$,
O. G. King$^5$,
 \and
R. Reeves$^6$
}

\affiliation{
$^{1}$National Radio Astronomy Observatory (NRAO), P.O. Box 0, Socorro, NM 87801, USA\\
email: {\tt wmax@nrao.edu} \\
$^{2}$Department of Physics, Purdue University, West Lafayette, IN 47907, USA\\
$^{3}$Aalto University Mets\"ahovi Radio Observatory, Mets\"ahovintie 114, 02540 Kylm\"al\"a, Finland\\
$^{4}$Department of Physics, University of Crete / Foundation for Research and Technology - Hellas, Heraklion 71003, Greece\\
$^{5}$Cahill Center for Astronomy and Astrophysics, California Institute of Technology, Pasadena, CA 91125, USA\\
$^{6}$Departamento de Astronom\'ia, Universidad de Concepci\'on, Casilla 160-C, Concepci\'on, Chile
}

\pubyear{2014}
\volume{313} 
\pagerange{xx--xx}
\jname{Extragalactic jets from every angle}
\editors{F. Massaro, C.C. Cheung, E. Lopez, A. Siemiginowska, eds.}
\begin{document}

\maketitle

\begin{abstract}
Since mid-2007 we have carried out a dedicated long-term monitoring programme at 15 GHz using the Owens Valley Radio Observatory 40 meter telescope (OVRO 40m). One of the main goals of this programme is to study the relation between the radio and gamma-ray emission in blazars and to use it as a tool to locate the site of high energy emission. Using this large sample of objects we are able to characterize the radio variability, and study the significance of correlations between the radio and gamma-ray bands. We find that the radio variability of many sources can be described using a simple power law power spectral density, and that when taking into account the red-noise characteristics of the light curves, cases with significant correlation are rare. We note that while significant correlations are found in few individual objects, radio variations are most often delayed with respect to the gamma-ray variations. This suggests that the gamma-ray emission originates upstream of the radio emission. Because strong flares in most known gamma-ray-loud blazars are infrequent, longer light curves are required to settle the issue of the strength of radio-gamma cross-correlations and establish confidently possible delays between the two. For this reason continuous multiwavelength monitoring over a longer time period is essential for statistical tests of jet emission models.
\end{abstract}

\firstsection % if your document starts with a section,

\section{Introduction}
Although the general picture of synchrotron emission at low energies and inverse Compton at high energies is well established, important aspects of blazars are not well understood. In particular, the location of the gamma-ray emission region is not clearly established. There are models and observations that argue for a location close to central engine \citep[$\ll 1$ pc, e.g.,][]{blandford+1995, tavecchio+2010}, and also another set of models and observations pointing to a location further down the jet, at tens of parsecs \citep[e.g.,][]{jorstad+2001, jorstad+2013}.

This problem would have a straightforward answer if imaging in gamma-rays was possible, but this is not the case with current technology where we can get angular resolutions of $\sim 0.1^{\circ}$ at the highest gamma-ray energies observable with the Large Area Telescope onboard the Fermi Gamma-ray Space Telescope \citep[\emph{Fermi}-LAT,][]{atwood+2009}. Nonetheless, we have a good idea about the origin of the radio emission, thanks to the submilliarcsecond resolution achievable with very long baseline interferometry \citep[e.g.,][]{kellermann+1998}. We also know that blazars are extremely variable at most wavelengths and one alternative is to use that property to put constraints on the relative location of the radio and gamma-ray emission. The basic idea of our programme is that if the radio and gamma-ray emission are produced in the same region, we expect to see correlated variability between the light curves of these two bands, while delayed emission will tell us about the relative location of these two emission regions. This observational programme requires simultaneous monitoring of a large sample of objects in the radio and gamma-ray bands. For the gamma-ray band we can use the excellent capabilities of \emph{Fermi}, that provide continuous monitoring of the sky with complete coverage every 3 hours. The complementary resources needed in the radio band are provided by the OVRO 40m monitoring programme described below.

\section{The OVRO 40 meter telescope blazar monitoring programme}

Since mid-2007 we have carried out a dedicated long-term monitoring programme at 15 GHz using the Owens Valley Radio Observatory 40 meter telescope. We are currently observing about 1800 blazars twice per week. The sample includes all the \emph{Fermi}-LAT detected blazars north of declination $-20^{\circ}$. A detailed description of the programme can be found in \citet{richards+2011}, and a list of publications and data in our website http://www.astro.caltech.edu/ovroblazars/.

One of the main advantages of continuously monitoring a large number of sources is that we can study the variability properties of a large sample, and their variation over different source populations. The results of such studies are presented in \citet{richards+2011} and \citet{richards+2014}. Similar studies including optical data are presented in \citet{hovatta+2014}. A question that is closely related to the existence of correlated radio/gamma-ray variability is the correlation of their mean fluxes. This question has been studied using single epoch surveys, but monitoring data can overcome several of the difficulties of those studies. Detailed discussions about this subject, and a look over different source classes is presented in \citet{ackermann+2011} and \citet{pavlidou+2012}. Besides those results, we discovered a 120-150 d quasi periodic oscillation in a blazar, the shortest one ever identified \citep[][]{king+2013}. We regularly contribute data for multiwavelength campaigns, which makes this programme a valuable result for the community.

\section{Characterization of the radio variability}

Variability is one the main characteristics of blazars, so we need to understand it in order to incorporate it in emission models. Efforts to incorporate this characteristic in models are starting to appear in the literature \citep[e.g., ][]{marscher2014, zhang+2014}, but they still require further development to provide testable observational predictions. A detailed understanding of the multiwavelegth variability properties of blazars is also essential for our understanding of the significance of multiwavelength cross-correlations as explained below. We can use the power spectral density (PSD), but there are other models that we do not consider here but that are discussed elsewhere and in this conference \citep[e.g.,][]{soboleska+2014}. Even for the simple PSD case, the characterization is complicated by the uneven sampling of the light curves.  We refer the reader to our recent paper \citet{max-moerbeck+2014b}, which describes the Monte Carlo approach used to for the PSD fit.

We characterize the variability using a simple power law PSD (PSD $\propto 1/f^{\beta}$). This model has a single parameter, the exponent $\beta$. Using 4 years of radio data on 1593 sources we can find good constraints for $\beta$ in 238 sources. We emphasize that the number of sources with good constraints is only a fraction of the total number of sources. This is because in most cases we need longer time series, or because the amplitude of the variability is not large enough to discriminate between different values of $\beta$. The lesson to learn is that getting errors for the PSD is as important as finding a best fit value, because it has a direct impact of the uncertainty in the estimation for the significance of a correlation as described in \citet{max-moerbeck+2014b} and briefly below. For the sources with constraints we found a range of values of $\beta$, with the property that they cluster around 2.3. Using our large sample we can look at variations of the PSD over different source populations (e.g., BL Lac versus FSRQ), but we did not find any variations of the PSD between classes, probably due to the large uncertainties in $\beta$ that we expect to reduce with longer radio light curves (Max-Moerbeck et al. in preparation).

\section{Significance of the cross-correlations}

With a characterization of the variability, we can look at the significance of cross-correlations. But before, we look at how important is the variability for the to study of the significance. Figure 12 of \citet{max-moerbeck+2014b} illustrates the different characteristics of the variability for time series with $\beta = 0,1$ and $2$. For $\beta = 0$ we see that no clear flares are seen in the light curve, but for $\beta = 2$ there appear clear flares similar to the ones observed in blazars. The problem arises when we look at correlations between those independent light curves. In this case \citep[see Figure 13 of][]{max-moerbeck+2014b}, we see that $\beta = 0$ does not produce large amplitude correlations, but that $\beta = 2$ produce large amplitude peaks in the cross-correlation. The lesson is that, whenever we correlate $1/f$ noise time series we will see correlations, so the most important question is how significant they are. To evaluate the significance of a cross-correlation we use Monte Carlo simulations. These simulation take into account the PSD, the sampling and also the observational noise. In Figure 28 of \citet{max-moerbeck+2014b} we can see how the results of the significance of correlations vary when we change the model used for the simulated light curves. This figure shows how important is to have a good characterization of the variability for the significance estimate. Having large errors in the PSD power-law exponent is equivalent to having fuzzy significance contours. Unfortunately this is the situation with most sources, and should be indicated when presenting cross-correlation significance results, as it one of the largest sources of uncertainty in these estimates.

\section{Results and their interpretation}

We studied the cross-correlation significance between 15 GHz and gamma-ray light curves. We used 4 years of radio data and 3 years of Fermi data for 86 sources \citep[][]{max-moerbeck+2014a}. Of those 86 sources only 41 are variable, have no long term trends and thus a significance analysis is possible. Of those 41 only 1 has a larger than $3\sigma$ significance and 2 more above $2.25\sigma$ significance, the level at which we expect to see one chance cross-correlation in a sample of 41. From this we find two main results: 

\begin{itemize}
\item
Only a minority of the sources have significant correlations
\item
In all the cases with correlations, the radio emission lags the gamma-ray emission.
\end{itemize}

These results demonstrate very clearly the difficulty of measuring statistically robust multiwavelength correlations and the care needed when comparing light curves even when many years of data are used.

We use a simple model to estimate the location of the gamma-ray emission site. In this model the emission is coming from the moving components we observe with VLBI. The peak of the gamma-ray flare occurs at $t_\gamma$, and some time later $t_r$, when the component propagates to optically thin part of the jet, beyond the radio core, we see the peak of the radio flare. Our cross-correlation result, gives $t_r - t_\gamma$, VLBA measurements and variability constraints provide the geometry of the jet and speed of the moving components.

We have enough ancillary data for two sources. For AO~0235+164 we consider the two peaks above $3\sigma$ seen in the correlation and we get a very weak constraint. For PKS~1502+106, the constraint is better and locates the gamma-ray emission a few parsecs away the central engine, but the errors are very large (22$\pm$15 pc for a conical jet, and 12$\pm$9 pc for the collimated jet case).

\section{KuPol: Polarization monitoring in the radio band}

The emission mechanism in the radio band is synchrotron radiation, and the variations can be due to changes in electron density or magnetic field, or to relativistic beaming effects. In this case polarization observations provide data on the evolution of the magnetic fields in the emission region. The OVRO 40 meter telescope has a completely new receiver, all the way from the front to the backend, that has been working since June 2014. This new receiver covers a broader band and will soon provide polarized flux density light curves.

\section{Summary}
This programme provides the largest data base of single frequency light curves and has produced many results by itself and for the community. We have used this data to learn about the variability of blazars in the radio band, and how it relates to the gamma-ray variations. Our main result is that only a minority of sources show significant correlations between the radio and gamma-ray bands, with radio lagging the gamma-ray emission.

\end{document}